\documentclass[twocolumn,showpacs,preprintnumbers,amsmath,amssymb]{revtex4}
\usepackage{graphicx}% Include figure files
\usepackage{dcolumn}% Align table columns on decimal point
\usepackage{bm}% bold math

\begin{document}

\preprint{APS/123-QED}

\title{Universal Time Tunneling}% Force line breaks with \\

\author{G\"unter Nimtz}
 \altaffiliation
 %[Also at ]%
 {G.Nimtz@uni-koeln.de}%Lines break automatically or can be forced with \\
%\author{Second Author}%
 %\email{Second.Author@institution.edu}
\affiliation{
II. Physikalisches Institut, Universt\"at zu K\"oln \\
Z\"ulpicherstrasse 77, 50937 K\"oln
}%

%\author{Charlie Author}
% \homepage{http://www.Second.institution.edu/~Charlie.Author}
%\affiliation{
%Second institution and/or address\\
%This line break forced% with \\
%}%

\date{\today}% It is always \today, today,
             %  but any date may be explicitly specified

\begin{abstract}

How much time does a tunneling wave packet spent in traversing a barrier?
Quantum mechanical calculations result in zero time inside a barrier .
In the nineties analogous tunneling experiments with microwaves were carried out. The results agreed with quantum mechanical calculations.
Electron tunneling time is hard to measure being extremely short and parasitic
effects due to the electric charge of electrons may be dominant. However, quite recently the atomic ionization tunneling time has been measured.
Experimental data of photonic, phononic, and electronic tunneling time is available now and will be presented. It appears that the tunneling time is a universal property independent of the field in question.
\end{abstract}

% PACS, the Physics and Astronomy
                             % Classification Scheme.
%\keywords{Suggested keywords}%Use showkeys class option if keyword
                              %display desired
\maketitle

\section{\label{sec:level1}Introduchtion }

\begin{figure}[htb]
\begin{center}
\includegraphics[width=0.3\textwidth]{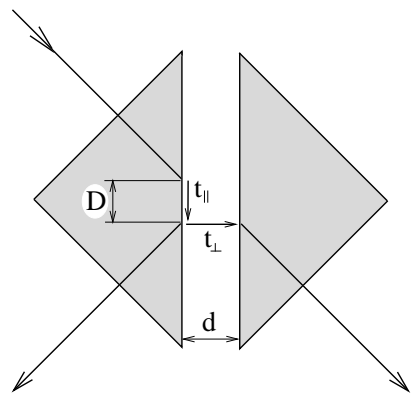}
\end{center}
\caption{Sketch of the time components in the investigated symmetrical beam design of double prisms under the condition of FTIR.
Reflected and transmitted signals were detected at the same time in spite
of the fact that the transmitted beam traveled an additional distance d  due to the gap \cite{Haibel2}.
\label{tcomponents}}
\end{figure}

Several theoretical investigations resulted in zero time tunneling, i.e. the time spent inside a barrier is zero~\cite{Hartman,Carniglia,Ali,Low,Wang}. Experimental studies with microwaves have shown that the measured short barrier traversal time is spent at the front boundary of a barrier, whereas actually zero time is spent inside the barrier~\cite{Nimtz}.  For example, as illustrated in Fig.\ref{tcomponents} the barrier traversal time was found to be zero in a frustrated total internal reflection (FTIR) experiment with a symmetrical double prisms set-up~\cite{Haibel2}.
The experimental results revealed for the traversal time $t_\bot$ = 0. Reflected and transmitted signals arrived the detector at the same time.
The measured short barrier traversal time is
of the order of magnitude of the reciprocal frequency T of the wave packets ~\cite{Haibel}. The time $\tau_\parallel$ arises at the barrier front boundary, as it is evident from inspecting Fig.~\ref{tcomponents}. This is followed by the fact that reflection and traversal time are equal.

\begin{eqnarray}
\tau & = &  t_\parallel  + t_\bot  =  t_\parallel \\
\tau & \approx & \frac{1}{\nu} = T ,
\end{eqnarray}
where $\tau$ is the measured barrier traversal time and $\nu$ the carrier frequency of the electromagnetic wave packet.

In the case of a wave packet with a rest mass the traversal time can be written by
\begin{eqnarray}
\tau \approx \frac{h}{E} ,
\end{eqnarray}
where h is the Planck constant and E is the particle energy.

This universal tunneling time behavior observation was found at the same time by Esposito studying theoretically the traversal time and obtained the modified relation
\begin{eqnarray}
\tau_A = \frac{1}{\nu} \cdot A, \label{A} \label{Espo}
\end{eqnarray}
where A is depending on the special barrier and wave packet in question~\cite{Esposito}.

Later experimental data on phonon and electron tunneling have pointed to a similar barrier traversal time~\cite{Yang,NimtzA}.
Quite recently, electronic tunneling data were measured in the ionization process of Helium \cite{Keller}, which also fit in this approximate
universal behavior of the traversal time of Eq.\ref{Espo}.

\begin{table*}
%\begin{sidewaystable}[htb]
\begin{center}
\begin{tabular}{|l|l|c|c|c|}

%\hline \textbf{Table: Tunneling time} &   &    &   &\\
\hline
\hline Tunneling barriers & Reference  & $\tau$   &  $T=1/\nu$ & $\tau_A$  \\
\hline
\hline {\it frustrated }   & Haibel/Nimtz & 117\,ps & 120\,ps & 81\,ps \\
\hline
\cline{2-4} {\it total reflection}   & Balcou/Dutriaux  &  30\,fs &  11.3\,fs & 36.8\,fs  \\
\hline
\cline{2-4} {\it  at double prisms}  & Mugnai et al. & 134\,ps & 100\,ps  & 87\,ps  \\
\cline{2-4}
\hline
\hline {\it photonic lattice} & Steinberg et al. & 2.13\,fs & 2.34\,fs & 2.02\,fs \\
\hline
\cline{2-4} &  Spielmann et al.  &  2.7\,fs  &  2.7\,fs & 2.98\,fs  \\
\hline
\hline {\it undersized waveguide} & Enders/Nimtz & 130\,ps & 115\,ps & 128\,ps \\
\hline
\hline {\it Ionization tunneling} & Eckle et al.  & 6\,as & 75\,as & 4.25\,as\\
\hline
\hline {\it acoustic (phonon) tunneling } & Yang et al. & 0.8\,$\mu$s &  1\,$\mu$s &  0.6$\mu$s \\
\hline
\hline
\end{tabular}
\end{center}
\caption{Tunneling Time}
\end{table*}
%\end{sidewaystable}
%\clearpage

\section{Experimental data}

Superluminal tunneling was observed first with
microwaves in undersized wave guides by Enders and Nimtz in 1992. Superluminal tunneling has been
reproduced at microwave and optical frequencies in photonic
band gap material, i.e. optical mirrors, and at double prisms  FTIR~\cite{Nimtz,Enders,Nimtz4}.

The barrier traversal time was obtained either by directly measuring the barrier traversal time or by calculating the time by the phase time approach
\begin{eqnarray}
\tau = - d\phi/d\omega,\label{phase}
\end{eqnarray}\\
where $\phi$ is the phase shift of the wave and $\omega$ is the
angular frequency; $\phi$ is given by the real part of the wave
number k times the distance x. In the case of evanescent modes
and of tunneling solutions  the wave number k is purely imaginary. Thus propagation of
evanescent or tunneling modes appears to take place in zero time \cite{Low}.

Zero time tunneling  of wave packets as
well as the barrier interaction time become most obvious in FTIR
with double prisms as studied in Refs.~\cite{Haibel2,NimtzA,Stahlhofen} for instance.  The optical double prisms were seen by Sommerfeld as the analog of quantum mechanical tunneling~\cite{Sommerfeld}. The symmetrical design for the paths of the beam  and the corresponding traveling time
components are sketched in Fig.\ref{tcomponents}. The experimental result revealed for the traversal time $t_\bot$ = 0, because of reflected and transmitted signals arrived the detectors at the same time. In FTIR there is a shift (coined Goos-H\"anchen shift) between the
incident and the reflected beam. The length of this shift amounts to about one wavelength
~\cite{Haibel2}. This shift along the boundary of the first prism
corresponds to the universal tunneling time of about one oscillation
time T ~\cite{Haibel,Esposito,NimtzA}. The shift represents the interaction
time at the boundary.

The approximate universal tunneling time appears to be valid also for sound
waves, having in mind that elastic and electromagnetic fields are described in an analogous mathematical approach.  Yang et al., for instance, investigated the tunneling time of ultrasonic pulses in the forbidden frequency gap of a phononic crystal. Ultrasonic pulses had a carrier frequency of 1 MHz. They measured a tunneling time between 0.6 and 1.0 $\mu s$~\cite{Yang}.

Even now, the extremely short atom ionization electron tunneling time was measured. Actually, this measured tunneling time does fit the modified universal tunneling time concept as shown in the Table. (It was assumed E = 54.39 eV and V$_0$ = 78.98 eV neglecting the applied laser field of about 10$^{-14}$ W/cm$^2$.)

\section{Discussion}
In 1962 Hartman's quantum mechanical study on tunneling of a wave packet revealed for
opaque barriers a superluminal velocity and a tunneling time independent of barrier length~\cite{Hartman,Nimtz5}. The calculations
were confirmed by the above mentioned photonic tunneling experiments, see for instance~Refs.\cite{Nimtz5,Nimtz2,Steinberg,Krausz}.
Haibel and Nimtz figured an empirical universal relation for the tunneling time out, which was observed with
several photonic barriers~\cite{Haibel}. The  empirical relation was theoretically substantiated by Esposito for both particles and evanescent optical modes~\cite{Esposito}. He has shown that the simple relation $\tau \approx \frac{1}{\nu} = T$ can be substituted by the general analytical expression of Eq.~\ref{A}
The factor A is given for the case of a tunneling particle described by the Schr\"odinger equation as follows,
\begin{eqnarray}
\tau_A = \frac{\hbar}{ \sqrt{E(V_0 - E)} }= \frac{1}{\nu} \cdot \frac{E}{4 \pi^2 (V_0 - E)},
\end{eqnarray}
where E and $V_0$ are the particle energy and the height of the square potential barrier, and E = h $\nu$.

In the case of FTIR at double prisms the factor becomes
\begin{eqnarray}
\tau_A = \frac{1}{\nu} \cdot \frac{n_1 \sin^2\theta}{\pi \cos\theta \sqrt{n_1^2 \sin^2\theta - n_2^2}} ,
\end{eqnarray}
where $n_1$ $\geq$ $n_2$ are the refractive indices of the prisms and of the gap, respectively. $\theta$ is the angle of incidence.
Schr\"odinger and Helmholtz equations result in analogous relations for the tunneling time. The factor corrects the $1/\nu = T$ value usually by some percent.

Tunneling modes propagate in zero time. This contradicts the often given
interpretation of tunneling by the uncertainty relation, where the particle
may acquire energy to overcome the barrier. However, this would cause a comparatively long vacuum barrier traversal time, much longer than the observed universal traversal time. The interaction with a square barrier takes place at the barrier's front and causes the short universal tunneling time presumably for all fields due to the same mathematical tunneling description.
\begin{acknowledgments}
We wish to acknowledge offering suggestions and encouragement in this field by Claus L\"ammerzahl.
\end{acknowledgments}

%\bibliography{apssamp}% Produces the bibliography via BibTeX.

\end{document}